\documentstyle[preprint,floats,pra,aps]{revtex}
\tightenlines
\begin{document}

\renewcommand{\thefootnote}{\fnsymbol{footnote}}

\def\pxb{\left(\p \times \B - \B \times \p \right)}

\def\rk{r_k}
\def\beq{\begin{equation}}
\def\eeq{\end{equation}}
\def\bea{\begin{eqnarray}}
\def\eea{\end{eqnarray}}
\def\nn{\nonumber}
\def\ba{\begin{array}}
\def\ea{\end{array}}
\def\0{{\mbox{\boldmath $0$}}}
\def\one{1\hskip -1mm{\rm l}}
\def\A{{\mbox{\boldmath $A$}}}
\def\B{{\mbox{\boldmath $B$}}}
\def\El{{\mbox{\boldmath $E$}}}
\def\F{{\mbox{\boldmath $F$}}}
\def\S{{\mbox{\boldmath $S$}}}
\def\P{{\mbox{\boldmath $P$}}}

\def\a{{\mbox{\boldmath $a$}}}
\def\p{{\mbox{\boldmath $p$}}}
\def\hatp{{\hat{\mbox{\boldmath $p$}}}}
\def\vpi{{\mbox{\boldmath $\pi$}}}
\def\hatvpi{\hat{\mbox{\boldmath $\pi$}}}
\def\r{{\mbox{\boldmath $r$}}}
\def\v{{\mbox{\boldmath $v$}}}
\def\w{{\mbox{\boldmath $w$}}}
\def\H{{\rm H}}
\def\hA{\hat{A}}
\def\hB{\hat{B}}
\def\i{{\rm i}}
\def\ih{\frac{\i}{\hbar}}
\def\ixh{\i \hbar}
\def\ddx{\frac{\partial}{\partial x}}
\def\ddy{\frac{\partial}{\partial y}}
\def\ddz{\frac{\partial}{\partial z}}
\def\ddt{\frac{\partial}{\partial t}}
\def\vsig{{\mbox{\boldmath $\sigma$}}}
\def\Al{{\mbox{\boldmath $\alpha$}}}
\def\ho{\hat{\cal H}_o}
\def\half{\frac{1}{2}}
\def\E{{\hat{\cal E}}}
\def\O{{\hat{\cal O}}}
\def\eps{\epsilon}
\def\g{\gamma}
\def\Vomeg{{\underline{\mbox{\boldmath $\Omega$}}}_s}
\def\hH{\hat{H}}
\def\Vsig{{\mbox{\boldmath $\Sigma$}}}
\def\Nab{{\mbox{\boldmath $\nabla$}}}
\def\curl{{\rm curl}}
\def\bh{\bar{H}}
\def\th{\tilde{H}}

\title
{\bf On the form of Lorentz-Stern-Gerlach force}

\author{Sameen Ahmed KHAN\thanks{khan@pd.infn.it,
~~~ http://www.pd.infn.it/$\sim$khan/}}
\author{Modesto PUSTERLA\thanks{pusterla@pd.infn.it,
~~~ http://www.pd.infn.it/$\sim$pusterla/}}
\address
{Dipartimento di Fisica Galileo Galilei
Universit\`{a} di Padova \\
Istituto Nazionale di Fisica Nucleare~(INFN) Sezione di Padova \\
Via Marzolo 8 Padova 35131 ITALY}
\maketitle
\begin{abstract}
In recent times there has been a renewed interest in the force 
experienced by a charged-particle with anomalous magnetic moment in the
presence of external fields. In this paper we address the basic question 
of the force experienced by a spin-$\frac{1}{2}$ point-like
charged-particle with magnetic and electric moments in the presence of
space-and time-dependent external electromagnetic fields, when derived
from the Dirac equation {\em via} the Foldy-Wouthuysen transformation
technique. It is interesting to note that the force thus derived differs 
from the ones obtained by various other prescriptions.
\end{abstract}

\medskip

\section{Introduction}
We present a derivation of the force experienced by a spin-$\frac{1}{2}$
point-like charged-particle with anomalous magnetic and anomalous electric
moments in the presence of space-and time-dependent external electromagnetic
fields, based {\em ab~initio} on the Dirac equation {\em via} the 
Foldy-Wouthuysen~(FW) transformation technique. In the present derivation
we neglect the radiation reaction and the electromagnetic fields are treated 
as classical.

In absence of spin the force experienced by a point-like charged-particle 
is completely described by the Lorentz force
law~($\F_L = q \left(\El + \v \times \B \right)$). In the regime where the
spin and the magnetic moment are to be taken into account the question of
the form of the force obtained from the relativistic quantum theory is 
still unresolved to this day, though extensive studies, using diverse 
approaches have been done since the discovery of quantum mechanics. This 
is evident from the numerous approaches/prescriptions which have been tried 
to address this basic question and are still being tried. Before proceeding
further we note that the expression quoted above constitutes the Lorentz
force.  The total force which we call as the Lorentz-Stern-Gerlach~(LSG)
force includes the Lorentz force as the basic constituent and all the 
other contributions coming from the spin, anomalous magnetic and electric
moments~{\it etc,}. The reason for this nomenclature will be clear as we 
proceed. 

Here we quote a few approaches which have been used to address the 
question of the force and acceleration experienced by a charged-particle.
A Lagrangian formalism based on the action principle has been
suggested~\cite{Nash-1}-\cite{Costella-1}. A Hamiltonian formalism is 
considered in~\cite{Chaichian-1}--\cite{Heinemann}. In the case of slowly 
varying electromagnetic fields an approach based on the Dirac equation 
{\em via} the WKB approximation scheme has been presented~\cite{Anandan-1}. 
In the context of the Aharonov-Bohm and Aharonov-Casher 
effects~\cite{Boyer}-\cite{Aharonov-1}, {\em the question as to whether 
neutron acceleration can occur in uniform electromagnetic fields is
also raised}~\cite{Russell}-\cite{Anandan-2}. In the very recent work of
Chaichian~\cite{Chaichian-1} it has been rightly pointed out that in the 
nonrelativistic limit the results of the above approaches do not coincide. 
This motivates us to examine the form of the force derived from the
Dirac equation using the FW-transformation~\cite{FW}-\cite{BD} scheme; 
note that the FW-transformation technique is the only one in which we can
take the meaningful nonrelativistic limit of the Dirac 
equation~\cite{Costella-2}. The FW-approach gives the expression for the
force in the presence of external time-dependent fields, the
nonrelativistic limit and a systematic procedure to obtain the
relativistic corrections to a desired degree of accuracy. In such a
derivation we also take into account the anomalous electric moment. We 
compare the results of our derivation with other approaches mentioned 
above. One should also note that a novel approach for producing polarized 
beams has been suggested using the Stern-Gerlach 
forces~\cite{CPP}-\cite{CJKP-1}.

\section{Section}
Let us consider the Dirac particle of rest mass $m_0$, charge $q$, 
anomalous magnetic moment $\mu_a$ and anomalous electric moment $\eps_a$. 
In presence of the external electromagnetic fields, the Dirac equation is
%01
\bea
\i \hbar
\frac{\partial}{\partial t} \left|\Psi_D \right\rangle 
= \hat{\rm H}_D \left|\Psi_D \right\rangle 
\eea
and the Dirac Hamiltonian ${\rm H}_D$, including the contributions of 
the anomalous magnetic moment and anomalous electric moment is given
by~\cite{Thaller}:
%02 03 04
\bea 
\hat{\rm H}_D & = & \beta m_0 c^2 + \E + \O \\
\E & = & + q \phi (\r) I - \mu_a \beta \Vsig \cdot \B 
+ \eps_a \beta \Vsig \cdot \El \\
\O & = & c \Al \cdot (- \ixh \Nab - q \A )
+ \i \mu_a \beta \Al \cdot \El + \i \eps_a \beta \Al \cdot \B \nn \\
\beta & = & \left(
\ba{cc}
\one & \0 \\
\0 & - \one
\ea \right)\,, \quad 
\Al =
\left(
\ba{cc}
\0 & \vsig \\
\vsig & \0
\ea \right)\,, \quad
\Vsig = 
\left(
\ba{cc}
\vsig & \0 \\
\0 & \vsig 
\ea \right)\,, \quad 
\label{Full-Dirac}
\eea
where $\vsig$ is the triplet of Pauli matrices.

In the nonrelativistic situation the upper pair of components of the 
Dirac spinor $\Psi_D$ are large compared to the lower pair of components.
The operator $\E$ which does not couple the large and small components of
$\Psi_D$ is called as {\em even} and $\O$ is called as an {\em odd} 
operator which couples the large to small components. Note that
$\beta \O = - \O \beta$ and $\beta \E = \E \beta$. 
This motivates us to look for a transformation which will eliminate the 
odd-part completely from the Dirac Hamiltonian. Such a transformation is 
available in the case of the free-particle. In the very general case of
time-dependent fields such a transformation is not known to exist. 
Therefore, one has to be content with an approximation procedure which 
reduces the strength of the odd-part to a desired degree of accuracy in 
powers of $\frac{1}{m_0 c^2}$. We shall follow the 
Foldy-Wouthuysen~\cite{FW}-\cite{BD} transformation technique to take the 
nonrelativistic limit of the Dirac Hamiltonian in~(\ref{Full-Dirac}) to 
reduce the strength of the odd-part in power series in $\frac{1}{m_0 c^2}$.
The result to the leading order, that is to order $\frac{1}{m_0 c^2}$ is 
formally given by
%05
\bea 
\i \hbar \ddt \left| \psi \right\rangle & = & \hat{\cal H}^{(2)} 
\left|\psi \right\rangle\,, \nn \\ 
\hat{\cal H}^{(2)} & = & 
m_0 c^2 \beta + \E + \frac{1}{2 m_0 c^2} \beta \O^2 
\label{FW-2}
\eea
and to next higher, order $\frac{1}{(m_0 c^2)^3}$ is given by
%06
\bea 
\i \hbar \ddt \left|\psi \right\rangle & = & \hat{\cal H}^{(4)} \left| 
\psi \right\rangle\,, \nn \\ 
\hat{\cal H}^{(4)} 
& = & 
m_0 c^2 \beta + \E + \frac{1}{2 m_0 c^2} \beta \O^2 \nn \\
& & - \frac{1}{8 m_0^2 c^4}
\left[\O , \left(\left[\O , \E \right] + \i \hbar \ddt \O \right) \right] 
- \frac{1}{\left(2 m_0 c^2\right)^3} \beta \O^4\,.
\label{FW-4}
\eea 
A detailed discussion of the FW-transformation and the derivation of the 
above Hamiltonians can be found in many places~(for instance the book by
Bjorken and Drell in~\cite{BD}).

As a first step we consider the case of a charged particle neglecting the
anomalous moments. In this case the Hamiltonian~(\ref{FW-4}) works out
to
%07
\bea
\hat{\cal H}^{(4)} & = & 
m_0 c^2 + q \phi 
+ \frac{1}{2 m_0}
\left(\hat{\pi}^2 - q \hbar \vsig \cdot \B \right) \nn \\
& & 
+ \frac{1}{8 m_0^2 c^2} \hbar q 
\vsig \cdot \left(\hatvpi \times \El - \El \times \hatvpi \right) \nn \\
& & 
- \frac{1}{8 m_0^3 c^2} 
\left\{
\hat{\pi}^4 + \hbar^2 q^2 B^2 
- \hbar q \left(\hat{\pi}^2 \left(\vsig \cdot \B \right) 
+ \left(\vsig \cdot \B \right) \hat{\pi}^2 \right)
\right\}
\label{H-four}
\eea
A detailed formula including the $\mu_a$ contibutions is given
by~(\ref{mu-H-four}) in the appendix.

\section{Section}
Now we use the Hamiltonian derived in~(\ref{H-four}) to compute 
the acceleration, $\a$ experienced by the particle using the Heisenberg 
representation,
%08
\beq 
\frac{d}{d t} \left\langle \O \right\rangle = 
\frac{\i}{\hbar} \left\langle \left[ \hat{\cal H} \,, \O \right] \right\rangle 
+ \left\langle \ddt \O \right\rangle\,.
\label{Heisenberg}
\eeq
and we omit the brackets $\langle \cdots \rangle$. Then we obtain,
%09
\bea
m_0 \dot{\r} & = & m_0 \frac{d}{d t} \r =
m_0 \frac{\i}{\hbar} \left[\hat{\cal H} \,, \r \right] \nn \\
& = &
\hatvpi - \frac{1}{4 m_0^2 c^2}
\left(\hat{\pi}^2 \hatvpi + \hatvpi \hat{\pi}^2 \right) %\nn \\ & &
- \frac{\hbar q}{4 m_0 c^2} \left(\vsig \times \El \right)
+ \frac{\hbar q}{4 m_0^2 c^2} 
\left(\hatvpi \left(\vsig \cdot \B \right)
+ \left(\vsig \cdot \B \right) \hatvpi \right)
\label{r-dot}
\eea
Using the above expression for $\dot{\r}$ we compute the acceleration
%10
\bea
m_0 \a & = & m_0 \frac{d }{d t} \dot{\r} = m_0 \ddot{\r} \nn \\
& = &
q \El - \frac{q}{4 m_0^2 c^2} 
\left(\hat{\pi}^2 \El + \El \hat{\pi}^2\right) \nn \\
& &
- \frac{q}{2 m_0}
\left(\hatvpi \times \B - \B \times \hatvpi \right) \nn \\
& & - \frac{q}{4 m_0^3 c^2}
\left(
\hat{\pi}^2 \left(\hatvpi \times \B - \B \times \hatvpi \right)
+
\left(\hatvpi \times \B - \B \times \hatvpi \right) \hat{\pi}^2 
\right) \nn \\
& &
- \frac{q}{4 m_0^2 c^2}
\left\{
\left(\hatvpi \cdot \El + \El \cdot \hatvpi\right) \hatvpi
+
\hatvpi \left(\hatvpi \cdot \El + \El \cdot \hatvpi\right) \right\} \nn \\
& &
+ \frac{\hbar q}{2 m_0} \Nab \left(\vsig \cdot \B \right) 
- \frac{\hbar q}{8 m_0^3 c^2} \left(
\hat{\pi}^2 \Nab \left(\vsig \cdot \B \right) 
+
\Nab \left(\vsig \cdot \B \right) \hat{\pi}^2 \right) \nn \\
& &
+ \frac{\hbar q^2}{4 m_0^2 c^2} 
\left(\El \left(\vsig \cdot \B \right) 
+ \left(\vsig \cdot \B \right) \El \right)
- \frac{\hbar q}{8 m_0^2 c^2}
\Nab \left(\vsig \cdot
\left(\hatvpi \times \El - \El \times \hatvpi \right) \right) \nn \\
& & 
+ \frac{\hbar q^2}{8 m_0^3 c^2} 
\left\{
\left(\hatvpi \times \B - \B \times \hatvpi \right) 
\left(\vsig \cdot \B \right) 
+
\left(\vsig \cdot \B \right) 
\left(\hatvpi \times \B - \B \times \hatvpi \right) 
\right\} \nn \\
& &
+ \frac{\hbar^2 q^2}{8 m_0^3 c^2} \Nab \left(B^2\right) 
+ \frac{\hbar q^2}{4 m_0^2 c^2} \El \left(\vsig \times \B \right) \nn \\
& &
- \frac{\hbar q}{4 m_0 c^2} \ddt \left(\vsig \times \B \right) 
+ \frac{\hbar q}{4 m_0^2 c^2} 
\left(
\hatvpi \ddt \left(\vsig \cdot \B \right) 
+ \ddt \left(\vsig \cdot \B \right) \hatvpi \right) \nn \\
& & + {\mbox{\boldmath $R$}}
\label{acceleration-4}
\eea
where the $\rk$-th component of ${\mbox{\boldmath $R$}}$ is
%11
\bea
\left({\mbox{\boldmath $R$}} \right)_{\rk} & = & 
- \frac{\hbar q}{8 m_0^2 c^2} 
\left\{ 
\hatvpi \cdot \Nab \left( \left(\vsig \times \El \right)_{\rk} \right)
+
\Nab \left( \left(\vsig \times \El \right)_{\rk} \right) \cdot \hatvpi
\right\} \nn \\
& & \qquad \qquad \rk = x\,, y\,, z\,, \quad
k = 1\,, 2\,, 3\,. 
\label{R-one}
\eea
The above expression for the acceleration can be related to the classical 
expression when we make the substitution 
$\frac{\hatvpi}{m_0} \longrightarrow {\mbox{\boldmath $v$}}$
where ${\mbox{\boldmath $v$}}$ is the velocity of the particle. With such 
a substitution and with 
$\beta = \frac{\left|\mbox{\boldmath $v$}\right|}{c}$
we get,
%12
\bea
{\mbox{\boldmath $a$}} & = &
\left(1 - \frac{1}{2} \beta^2 \right) 
\frac{q}{m_0} \left(\El + {\mbox{\boldmath $v$}} \times \B \right)
- \frac{q}{m_0} \frac{{\mbox{\boldmath $v$}}}{c^2}
\left({\mbox{\boldmath $v$}} \cdot \El \right) \nn \\
& &
\left(1 - \frac{1}{2} \beta^2 \right) 
\frac{\hbar q}{2 m_0} \Nab \left(\vsig \cdot \B\right) 
- 
\frac{\hbar q}{4 m_0 c^2} \Nab 
\left(\vsig \cdot \left(\v \times \El \right) \right) \nn \\
& &
\frac{1}{m_0 c^2} 
\left\{q \left(\El + {\mbox{\boldmath $v$}} \times \B \right)\right\}
\frac{\hbar q}{2 m_0} \left(\vsig \cdot \B\right) \nn \\
& & + \cdots
\label{acceleration-limit}
\eea
The above derivation is consistent with the result~\cite{Landau} of
classical electrodynamics
%13
\bea
{\mbox{\boldmath $a$}} & = &
\frac{q}{m_0} \sqrt{1 - \beta^2}
\left\{\El + {\mbox{\boldmath $v$}} \times \B 
- \frac{{\mbox{\boldmath $v$}}}{c^2}
\left({\mbox{\boldmath $v$}} \cdot \El \right) \right\}
\label{a-classical}
\eea

In the nonrelativistic limit the force ${\mbox{\boldmath $F$}}$ is 
well-approximated by the expression
${\mbox{\boldmath $F$}} = m_0 {\mbox{\boldmath $a$}}$. So we can use the
expression for the acceleration, ${\mbox{\boldmath $a$}}$ derived using 
the Foldy-Wouthuysen technique to express the force experienced by the
charged particle.

The leading order Foldy-Wouthuysen Hamiltonian, when we take the anomalous 
magnetic moment and anomalous electric moment into account is given by
%14  15
\bea
\hat{\cal H}^{(2)} & = & 
m_0 c^2 - \mu_a \vsig \cdot \B 
+ \eps_a \vsig \cdot \El + q \phi \nn \\
& & 
+ \frac{1}{2 m_0 c^2}
\left\{
c^2 \left(\hat{\pi}^2 - q \hbar \vsig \cdot \B \right)
+ \left(\mu_a \El + \epsilon_a \B \right)^2
\phantom{\frac{\partial}{\partial t}} \right. \nn \\
& & \qquad \quad \left.
\phantom{\frac{\partial}{\partial t}}
+ \mu_a c \vsig \cdot \left(\hatvpi \times \El - \El \times \hatvpi \right)
+ \epsilon_a c \vsig \cdot \left(\hatvpi \times \B - \B \times \hatvpi \right) 
\right\}
\label{H-two}
\eea
Next, to leading order Hamiltonian is given in~(\ref{mu-H-four}) of the
appendix.

Now we use the above derived Hamiltonians in~(\ref{H-two}) to compute the
Lorentz-Stern-Gerlach force and we get 
%15 16
\bea
\dot{\r} & = &
\frac{d}{d t} \r = \frac{1}{m_0} 
\left\{\hatvpi 
- \left(\frac{\mu_a}{c} \left(\vsig \times \El \right)
+
\frac{\epsilon_a}{c} \left(\vsig \times \B \right) \right)
\right\} \nn \\
& = & 
\frac{1}{m_0}
\hat{\mbox{\boldmath $\prod$}}
\label{kinetic}
\eea
Where $\hat{\mbox{\boldmath $\prod$}}$ is the {\em kinetic momentum}.
The Lorentz-Stern-Gerlach force in the absence of $\epsilon_a$ is:
%17
\bea
{\mbox{\boldmath $F$}} & = &
\frac{d }{d t} \hat{\mbox{\boldmath $\prod$}}
=
\frac{\i}{\hbar} \left[\hat{\cal H} \,, 
\hat{\mbox{\boldmath $\prod$}} \right]
+
\ddt \hat{\mbox{\boldmath $\prod$}} \nn \\
& = & 
q \left\{\El + \frac{1}{2 m_0} 
\left(\hatvpi \times \B - \B \times \hatvpi \right) \right\} \nn \\
& & 
+ \left(\mu_a + \frac{q \hbar}{2 m_0} \right)
\Nab \left(\vsig \cdot \B \right) 
- \frac{\mu_a}{c} \ddt \left(\vsig \times \El \right) \nn \\
& &
- \frac{\mu_a}{2 m_0 c} 
\Nab
\left(\vsig \cdot
\left(\hatvpi \times \El - \El \times \hatvpi \right)\right) \nn \\
& & 
+ 2 \frac{\mu_a}{\hbar c} \left(\mu_a + \frac{\hbar q}{2 m_0}\right)
\El \times \left(\vsig \times \B \right)
- \frac{1}{2 m_0 c^2} \Nab \left(\mu_a^2 \El^2 \right) \nn \\
& &
+ \i \frac{\mu_a^2}{2 m_0 c^2 \hbar} 
\left(\left(\hatvpi \times \El \right) \times \El
- \El \times \left(\El \times \hatvpi\right) \right) \nn \\
& &
+
\frac{\mu_a^2}{2 m_0 c^2 \hbar} 
\left\{
\left(\vsig \times
\left(\hatvpi \times \El - \El \times \hatvpi \right)
\right) \times \El 
-
\El \times
\left(\vsig \times
\left(\hatvpi \times \El - \El \times \hatvpi \right) \right)
\right\} \nn \\
& &
+ {\mbox{\boldmath $R$}}
\label{mu-two}
\eea
where the $\rk$-th component of ${\mbox{\boldmath $R$}}$ is
%17 18
\bea
\left({\mbox{\boldmath $R$}} \right)_{\rk} & = &
-
\frac{\mu_a}{2 m_0 c}
\left\{ 
\hatvpi \cdot \Nab \left( \left(\vsig \times \El \right)_{\rk} \right)
+
\Nab \left( \left(\vsig \times \El \right)_{\rk} \right) \cdot \hatvpi
\right\} \nn \\
& & \qquad \qquad \rk = x\,, y\,, z\,, \quad
k = 1\,, 2\,, 3\,. 
\label{R-two}
\eea

For simplicity we first consider the the acceleration~(or equivalently the 
force) experienced by a neutron
%18
\bea
{\mbox{\boldmath $F$}}_{neutron} & = &
+ \mu_a \Nab \left(\vsig \cdot \B \right) 
- \frac{\mu_a}{c} \ddt \left(\vsig \times \El \right) \nn \\
& &
- \frac{\mu_a}{2 m_0 c} 
\Nab
\left(\vsig \cdot
\left(\hatp \times \El - \El \times \hatp \right)\right) \nn \\
& & 
+ 2 \frac{\mu_a^2}{\hbar c}
\El \times \left(\vsig \times \B \right)
- \frac{1}{2 m_0 c^2} \Nab \left(\mu_a^2 \El^2 \right) \nn \\
& &
+ \i \frac{\mu_a^2}{2 m_0 c^2 \hbar} 
\left(\left(\hatp \times \El \right) \times \El
- \El \times \left(\El \times \hatp\right) \right) \nn \\
& &
+
\frac{\mu_a^2}{2 m_0 c^2 \hbar} 
\left\{
\left(\vsig \times
\left(\hatp \times \El - \El \times \hatp \right)
\right) \times \El 
-
\El \times
\left(\vsig \times
\left(\hatp \times \El - \El \times \hatp\right)\right)
\right\} \nn \\
& & + \cdots \nn \\
& & + {\sf O} \left(\mu_a^3\right)
\label{f-neutron-1}
\eea
In the above expression in~(\ref{f-neutron-1}) the leading terms have 
been retained and the~``$\cdots$'' indicates the higher order terms. The
complete expression is given in~(\ref{f-neutron-a}) in the appendix. The 
detailed formulae shall be given in an appendix at the end. This is the 
case where ever the~``$\cdots$'' appear in the expressions.

From the expression~(\ref{f-neutron-1}) we conclude that the leading 
order~(linear in $\mu_a$) contributions to the neutron acceleration come 
through the gradients and the time derivatives of the electromagnetic 
fields. Such contributions disappear in the case of uniform and constant 
fields respectively. The next-to-leading order contributions come from 
the terms of the type $\mu_a^2 \El \times \left(\vsig \times \B\right)$. 
Such contributions do not vanish and hence we have neutron acceleration 
even in the presense of uniform fields. Such accelerations are 
quadratic~(and higher powers) in~$\mu_a$ and hence are very small.

In the expression~(\ref{mu-two}) for the Lorentz-Stern-Gerlach force if
we substitute $\mu_a = g \frac{\hbar |q|}{4 m_0}$ and $q = - e$ we get 
the often mentioned term,
$+ g \left(g - 2 \right) \El \times \left(\vsig \times \B \right)$
in~\cite{Anandan-1}.

In the presence of the anomalous electric moment~$\epsilon_a$ the 
Lorentz-Stern-Gerlach force is:
%FULL-i
%20 21
\bea
{\mbox{\boldmath $F$}} & = &
q \left\{\El + 
\frac{1}{2 m_0} 
\left(\hatvpi \times \B - \B \times \hatvpi \right) \right\} \nn \\
& & 
+ 
\left(\mu_a + \frac{q \hbar}{2 m} \right)
\Nab \left(\vsig \cdot \B \right)
-
\eps_a \Nab \left(\vsig \cdot \El \right) \nn \\
& &
- 
\frac{1}{c} 
\ddt \left(
\mu_a \left(\vsig \times \El \right)
+
\eps_a \left(\vsig \times \B \right) \right) \nn \\
& &
- \frac{\mu_a}{2 m_0 c} 
\Nab \left(\vsig \cdot
\left(\hatvpi \times \El - \El \times \hatvpi \right)\right)
- \frac{\epsilon_a}{2 m_0 c} 
\Nab \left(\vsig \cdot
\left(\hatvpi \times \B - \B \times \hatvpi \right)\right) \nn \\
& &
\cdots
\label{f-full-i}
\eea

\section{Conclusions and Summary}

As can be seen above we get a variety of terms contributing to the 
Lorentz-Stern-Gerlach force. 

The nonrelativistic static limit coincides with the usual ``classical''
formula if $\B$ is time-independent, inhomogeneous and $\El$ is absent in
the lab system. Otherwise there are differences even at low non-relativistic
velocities. In particular one may consider the force terms
$\frac{\mu_a}{2 m_0 c} \Nab \left(\vsig \cdot \left({\mbox{\boldmath $v$}} 
\times \El \right) \right)$
and 
$\frac{1}{2 m_0 c^2} \Nab \left(\mu_a^2 \El^2 \right)$
that are present whenever a spin-$\frac{1}{2}$ particle with charge
enters into an inhomogeneous static electric field~(in absence of $\B$). 

Another relevant point to be noted is the force experienced by a 
neutron~(more generally by an electrically neutral particle). In this case 
we find contributions  even when the fields are homogeneous and static.

The LSG force derived  using the FW-tchnique differs from the other 
approaches which use a ``classical'' or ``semiclassical'' treatment of the 
relativistic Stern-Gerlach force~\cite{Heinemann}.

Only experiments with very high precision can conclude about the finer 
differences in the various expressions for the force.

\bigskip

\noindent
{\bf Acknowledgement}\\
The authors are very greatful to Prof. R.~Jagannathan~(Institute of 
Mathematical Science, Madras, India) for very useful discussions on the
subjects dealt in this paper.

\renewcommand{\theequation}{A{\arabic{equation}}}
\setcounter{equation}{0}

\begin{center}
{\bf APPENDIX}
\end{center}

For the general case including, the contributions of the anomalous
magnetic moment, the Hamiltonian~(in~\ref{FW-4}) works out to
%A01  14
\bea
\hat{\cal H}^{(4)} & = & 
m_0 c^2 + q \phi + \frac{1}{2 m_0} \hat{\pi}^2 \nn \\
& &
- \left(\frac{\hbar q}{2 m_0} + \mu_a \right) \left(\vsig \cdot \B \right)
+
\frac{\mu_a}{2 m_0 c}
\vsig \cdot \left(\hatvpi \times \El - \El \times \hatvpi \right)
+
\frac{\mu_a^2}{2 m_0 c^2} E^2 \nn \\
& &
+ 
\frac{1}{8 m_0^2 c^4}
\left\{
+ \hbar q c^2 
\vsig \cdot \left(\hatvpi \times \El - \El \times \hatvpi \right)
+ 2 \mu_a \hbar q c E^2 \right. \nn \\
& & \qquad \qquad
- \mu_a^2 \left(
\left(\vsig \cdot \hatvpi \right)
\left(\hatvpi \cdot \B + \B \cdot \hatvpi \right)
+
\left(\hatvpi \cdot \B + \B \cdot \hatvpi \right)
\left(\vsig \cdot \hatvpi \right) \right) \nn \\
\phantom{\frac{\partial}{\partial t}}
& & \qquad \qquad \left.
- \mu_a^2 \hbar c 
\vsig \cdot \left(\Nab \left(\El \cdot \B + \B \cdot \El \right)\right)
+
4 \mu_a^3 \left(\vsig \cdot \El \right) \left(\El \cdot \B\right)
\right\} \nn \\
& & 
- \frac{1}{\left(2 m_0 c^2 \right)^3}
\left\{ 
c^4 \left(
\hat{\pi}^4 + \hbar^2 q^2 B^2
- \hbar q \left(\hat{\pi}^2 \left(\vsig \cdot \B \right) 
+ \left(\vsig \cdot \B \right) \hat{\pi}^2 \right) \right) \right. \nn \\
& & \qquad \qquad \quad
+ \mu_a c^3 \left(\hat{\pi}^2 
\vsig \cdot \left(\hatvpi \times \El - \El \times \hatvpi \right)
+
\vsig \cdot \left(\hatvpi \times \El - \El \times \hatvpi \right)
\hat{\pi}^2 \right) \nn \\
& & \qquad \qquad \quad
- \mu_a c \hbar q \left(
\B \cdot \left(\hatvpi \times \El - \El \times \hatvpi \right)
+ 
\left(\hatvpi \times \El - \El \times \hatvpi \right) \cdot \B
\right) \nn \\
& & \qquad \qquad \quad
- \i \mu_a c \hbar q \vsig \cdot \left(
\B \times \left(\hatvpi \times \El - \El \times \hatvpi \right)
+ 
\left(\hatvpi \times \El - \El \times \hatvpi \right) \times \B \right) \nn \\
& & \qquad \qquad \quad
+ \mu_a^2 c^2 \left(
\left(\hat{\pi}^2 E^2 + E^2 \hat{\pi}^2 \right) 
+
\left(\hatvpi \times \El - \El \times \hatvpi \right)^2 \right) \nn \\
& & \qquad \qquad \quad
+ \i \mu_a^2 c^2 \vsig \cdot \left(
\left(\hatvpi \times \El - \El \times \hatvpi \right)
\times
\left(\hatvpi \times \El - \El \times \hatvpi \right) \right) \nn \\
& & \qquad \qquad \quad
- \mu_a^2 c^2 \hbar q \left(E^2 \left(\vsig \cdot \B \right) +
\left(\vsig \cdot \B\right) E^2 \right) \nn \\
& & \qquad \qquad \quad
+ \mu_a^3 c 
\left(
E^2 \vsig \cdot 
\left(\hatvpi \times \El - \El \times \hatvpi \right) 
+ 
\vsig \cdot
\left(\hatvpi \times \El - \El \times \hatvpi \right) E^2 \right) \nn \\
& & \qquad \qquad \quad \left.
+ \mu_a^4 E^4 
\right\}\,.
\label{mu-H-four}
\eea
The total acceleration~(or equivalently the force) experienced by a neutron
when both $\mu_a$ and $\epsilon_a$ are taken into account is:
%A02 19
\bea
{\mbox{\boldmath $F$}} & = &
\mu_a \Nab \left(\vsig \cdot \B \right)
-
\eps_a \Nab \left(\vsig \cdot \El \right) \nn \\
& &
- 
\frac{1}{c} 
\ddt \left(
\mu_a \left(\vsig \times \El \right)
+
\eps_a \left(\vsig \times \B \right) \right) \nn \\
& &
- \frac{\mu_a}{2 m_0 c} 
\Nab \left(\vsig \cdot
\left(\hatp \times \El - \El \times \hatp \right)\right)
- \frac{\epsilon_a}{2 m_0 c} 
\Nab \left(\vsig \cdot
\left(\hatp \times \B - \B \times \hatp \right)\right) \nn \\
& &
-
\frac{1}{2 m_0 c^2} 
\Nab \left\{
\mu_a^2 \El^2 + \eps_a^2 \B^2 
+ \mu_a \eps_a \left(\El \cdot \B + \B \cdot \El \right)
\right\} \nn \\
& &
-
2 \frac{\mu_a^2}{\hbar c}
\left(\vsig \times \B \right) \times \El 
-
2 \frac{\eps_a^2}{\hbar c}
\left(\vsig \times \El \right) \times \B \nn \\
& &
+
2 \frac{\mu_a \eps_a}{\hbar c}
\left(
\left(\vsig \times \El \right) \times \El
-
\left(\vsig \times \B \right) \times \B
\right) \nn \\
& &
-
\i \frac{\mu_a^2}{2 m_0 c^2 \hbar} 
\left(
\left(\hatp \times \El \right) \times \El
-
\El \times \left(\El \times \hatp\right)
\right) \nn \\
& &
-
\i \frac{\eps_a^2}{2 m_0 c^2 \hbar} 
\left(
\left(\hatp \times \B \right) \times \B
-
\B \times \left(\B \times \hatp\right)
\right) \nn \\
& &
+
\frac{\mu_a^2}{2 m_0 c^2 \hbar} 
\left\{
\left(\vsig \times
\left(\hatp \times \El - \El \times \hatp \right)
\right) \times \El 
-
\El \times
\left(\vsig \times
\left(\hatp \times \El - \El \times \hatp \right) \right) 
\right\} \nn \\
& &
+
\frac{\eps_a^2}{2 m_0 c^2 \hbar} 
\left\{
\left(\vsig \times
\left(\hatp \times \B - \B \times \hatp \right)
\right) \times \B
-
\B \times
\left(\vsig \times
\left(\hatp \times \B - \B \times \hatp \right) 
\right) 
\right\} \nn \\
& &
-
\i 
\frac{\mu_a \eps_a}{2 m_0 c^2 \hbar} 
\left\{
\left(\hatp \times \B - \B \times \hatp \right) \times \El
+
\El \times \left(\hatp \times \B - \B \times \hatp \right) 
\phantom{\frac{\mu_a \eps_a^2}{2 m_0 c^2 \hbar}} \right. \nn \\
& & \quad \quad \left. 
\phantom{\frac{\mu_a \eps_a^2}{2 m_0 c^2 \hbar}}
+ \left(\hatp \times \El - \El \times \hatp \right) \times \B
+
\B \times \left(\hatp \times \El - \El \times \hatp \right) 
\right\} \nn \\
& &
+
\frac{\mu_a \eps_a}{2 m_0 c^2 \hbar} 
\left\{
\left(\vsig \times
\left(\hatp \times \B - \B \times \hatp \right)
\right) \times \El
-
\El \times 
\left(\vsig \times
\left(\hatp \times \B - \B \times \hatp \right)
\right)
\phantom{\frac{\mu_a \eps_a}{2 m_0 c^2 \hbar}} \right. \nn \\
& & \quad \quad \left.
\phantom{\frac{\mu_a \eps_a^2}{2 m_0 c^2 \hbar}}
+ \left(\vsig \times
\left(\hatp \times \El - \El \times \hatp \right)
\right) \times \B
-
\B \times 
\left(\vsig \times
\left(\hatp \times \El - \El \times \hatp \right)
\right) 
\right\} \nn \\ 
& &
+ {\mbox{\boldmath $R$}} \nn \\
\label{f-neutron-a}
\eea
where the $\rk$-th component of ${\mbox{\boldmath $R$}}$ is
%
%A03 22.2
\bea
\left({\mbox{\boldmath $R$}} \right)_{\rk} & = & 
-
\frac{\mu_a}{2 m_0 c} 
\left\{ 
\hatp \cdot \Nab \left(\left(\vsig \times \El \right)_{\rk} \right)
+
\Nab \left( \left(\vsig \times \El \right)_{\rk} \right) \cdot \hatp
\right\} \nn \\
& &
-
\frac{\eps_a}{2 m_0 c} 
\left\{ 
\hatp \cdot \Nab \left( \left(\vsig \times \B \right)_{\rk} \right)
+
\Nab \left( \left(\vsig \times \B \right)_{\rk} \right) \cdot \hatp
\right\} \nn \\
& & \qquad \qquad \rk = x\,, y\,, z\,, \quad
k = 1\,, 2\,, 3\,. 
\label{R-three-a}
\eea

In the presence of the anomalous electric moment~$\epsilon_a$ the 
Lorentz-Stern-Gerlach force is:
%FULL
%A04 21
\bea
{\mbox{\boldmath $F$}} & = &
q \left\{\El + 
\frac{1}{2 m_0} 
\left(\hatvpi \times \B - \B \times \hatvpi \right) \right\} \nn \\
& & 
+ 
\left(\mu_a + \frac{q \hbar}{2 m} \right)
\Nab \left(\vsig \cdot \B \right)
-
\eps_a \Nab \left(\vsig \cdot \El \right) \nn \\
& &
- 
\frac{1}{c} 
\ddt \left(
\mu_a \left(\vsig \times \El \right)
+
\eps_a \left(\vsig \times \B \right) \right) \nn \\
& &
- \frac{\mu_a}{2 m_0 c} 
\Nab \left(\vsig \cdot
\left(\hatvpi \times \El - \El \times \hatvpi \right)\right)
- \frac{\epsilon_a}{2 m_0 c} 
\Nab \left(\vsig \cdot
\left(\hatvpi \times \B - \B \times \hatvpi \right)\right) \nn \\
& &
-
\frac{1}{2 m_0 c^2} 
\Nab \left\{
\mu_a^2 \El^2 + \eps_a^2 \B^2 
+ \mu_a \eps_a \left(\El \cdot \B + \B \cdot \El \right)
\right\} \nn \\
& &
+
\frac{\mu_a q}{m_0 c} \left(
\left(\vsig \times \El \right) \times \B 
- 
\left(\vsig \times \B \right) \times \El \right) \nn \\
& &
-
2 \frac{\mu_a^2}{\hbar c}
\left(\vsig \times \B \right) \times \El 
-
2 \frac{\eps_a^2}{\hbar c}
\left(\vsig \times \El \right) \times \B \nn \\
& &
+
2 \frac{\mu_a \eps_a}{\hbar c}
\left(
\left(\vsig \times \El \right) \times \El
-
\left(\vsig \times \B \right) \times \B
\right) \nn \\
& &
-
\i \frac{\mu_a^2}{2 m_0 c^2 \hbar} 
\left(
\left(\hatvpi \times \El \right) \times \El
-
\El \times \left(\El \times \hatvpi\right)
\right) \nn \\
& &
-
\i \frac{\eps_a^2}{2 m_0 c^2 \hbar} 
\left(
\left(\hatvpi \times \B \right) \times \B
-
\B \times \left(\B \times \hatvpi\right)
\right) \nn \\
& &
+
\frac{\mu_a^2}{2 m_0 c^2 \hbar} 
\left\{
\left(\vsig \times
\left(\hatvpi \times \El - \El \times \hatvpi \right)
\right) \times \El 
-
\El \times
\left(\vsig \times
\left(\hatvpi \times \El - \El \times \hatvpi \right) \right) 
\right\} \nn \\
& &
+
\frac{\eps_a^2}{2 m_0 c^2 \hbar} 
\left\{
\left(\vsig \times
\left(\hatvpi \times \B - \B \times \hatvpi \right)
\right) \times \B
-
\B \times
\left(\vsig \times
\left(\hatvpi \times \B - \B \times \hatvpi \right) 
\right) 
\right\} \nn \\
& &
-
\i 
\frac{\mu_a \eps_a}{2 m_0 c^2 \hbar} 
\left\{
\left(\hatvpi \times \B - \B \times \hatvpi \right) \times \El
+
\El \times \left(\hatvpi \times \B - \B \times \hatvpi \right) 
\phantom{\frac{\mu_a \eps_a^2}{2 m_0 c^2 \hbar}} \right. \nn \\
& & \quad \quad \left. 
\phantom{\frac{\mu_a \eps_a^2}{2 m_0 c^2 \hbar}}
+ \left(\hatvpi \times \El - \El \times \hatvpi \right) \times \B
+
\B \times \left(\hatvpi \times \El - \El \times \hatvpi \right) 
\right\} \nn \\
& &
+
\frac{\mu_a \eps_a}{2 m_0 c^2 \hbar} 
\left\{
\left(\vsig \times
\left(\hatvpi \times \B - \B \times \hatvpi \right)
\right) \times \El
-
\El \times 
\left(\vsig \times
\left(\hatvpi \times \B - \B \times \hatvpi \right)
\right)
\phantom{\frac{\mu_a \eps_a}{2 m_0 c^2 \hbar}} \right. \nn \\
& & \quad \quad \left.
\phantom{\frac{\mu_a \eps_a^2}{2 m_0 c^2 \hbar}}
+ \left(\vsig \times
\left(\hatvpi \times \El - \El \times \hatvpi \right)
\right) \times \B
-
\B \times 
\left(\vsig \times
\left(\hatvpi \times \El - \El \times \hatvpi \right)
\right) 
\right\} \nn \\ 
& & 
+
{\mbox{\boldmath $R$}}
\label{f-full}
\eea
where the $\rk$-th component of ${\mbox{\boldmath $R$}}$ is
%A05 22
\bea
\left({\mbox{\boldmath $R$}} \right)_{\rk} & = & 
-
\frac{\mu_a}{2 m_0 c} 
\left\{ 
\hatvpi \cdot \Nab \left(\left(\vsig \times \El \right)_{\rk} \right)
+
\Nab \left( \left(\vsig \times \El \right)_{\rk} \right) \cdot \hatvpi
\right\} \nn \\
& &
-
\frac{\eps_a}{2 m_0 c} 
\left\{ 
\hatvpi \cdot \Nab \left( \left(\vsig \times \B \right)_{\rk} \right)
+
\Nab \left( \left(\vsig \times \B \right)_{\rk} \right) \cdot \hatvpi
\right\} \nn \\
& & \qquad \qquad \rk = x\,, y\,, z\,, \quad
k = 1\,, 2\,, 3\,. 
\label{R-three-f}
\eea

As can be seen above we get a variety of terms contributing to the 
Lorentz-Stern-Gerlach force. 

%BIBLIOGRAPHY

\end{document}